\documentclass{aastex631}
\usepackage{amsmath}
\usepackage{soul}
\usepackage{graphicx}
\usepackage{subfigure}
\usepackage{fourier}
\usepackage{arcs}
\usepackage{url}
\usepackage{color}
\usepackage{amsbsy}
\usepackage{mathrsfs}
\usepackage{bm}
\usepackage{booktabs}
\usepackage{multirow}
\usepackage{url}
\usepackage{ulem}
\usepackage{makecell}
\usepackage{media9}
\usepackage{natbib}
\usepackage{tikz}
\begin{document}

\title{MHD modeling of magnetic flux evolution around solar maximum by the coronal model COCONUT}

\author[0000-0002-4217-6990]{Hao P. Wang}
\affiliation{Centre for Mathematical Plasma-Astrophysics, Department of Mathematics, KU Leuven, Celestijnenlaan 200B, 3001 Leuven, Belgium; \url{Stefaan.Poedts@kuleuven.be}; \url{andrea.lani@kuleuven.be}; \url{haopeng.wang1@kuleuven.be}}

\author[0000-0002-1743-0651]{Stefaan Poedts}
\affiliation{Centre for Mathematical Plasma-Astrophysics, Department of Mathematics, KU Leuven, Celestijnenlaan 200B, 3001 Leuven, Belgium}
\affiliation{Institute of Physics, University of Maria Curie-Skłodowska, ul. Radziszewskiego 10, 20-031 Lublin, Poland}

\author[0000-0003-4017-215X]{Andrea Lani}
\affiliation{Centre for Mathematical Plasma-Astrophysics, Department of Mathematics, KU Leuven, Celestijnenlaan 200B, 3001 Leuven, Belgium}

\author[0009-0006-7051-0438]{Jun Y. Liu}
\affiliation{Centre for Mathematical Plasma-Astrophysics, Department of Mathematics, KU Leuven, Celestijnenlaan 200B, 3001 Leuven, Belgium}
\affiliation{CAS Key Laboratory of Geospace Environment, Department of Geophysics and Planetary Sciences, University of Science and Technology of China, Hefei, 230026, People's Republic of China}

\author[0000-0002-7422-1127]{Quentin Noraz}
\affiliation{Centre for Mathematical Plasma-Astrophysics, Department of Mathematics, KU Leuven, Celestijnenlaan 200B, 3001 Leuven, Belgium}

\author[0000-0002-4014-1815]{Luis Linan}
\affiliation{Centre for Mathematical Plasma-Astrophysics, Department of Mathematics, KU Leuven, Celestijnenlaan 200B, 3001 Leuven, Belgium}

\author[0000-0002-1986-4496]{Tinatin Baratashvili}
\affiliation{Centre for Mathematical Plasma-Astrophysics, Department of Mathematics, KU Leuven, Celestijnenlaan 200B, 3001 Leuven, Belgium}

\author[0000-0003-4616-947X]{Hyun-Jin Jeong}
\affiliation{Centre for Mathematical Plasma-Astrophysics, Department of Mathematics, KU Leuven, Celestijnenlaan 200B, 3001 Leuven, Belgium}
\affiliation{School of Space Research, Kyung Hee University, Yongin, 17104, Republic of Korea}

\author[0000-0003-3670-4678]{Rayan Dhib}
\affiliation{Centre for Mathematical Plasma-Astrophysics, Department of Mathematics, KU Leuven, Celestijnenlaan 200B, 3001 Leuven, Belgium}

\author[0000-0001-8495-9179]{Wenwen Wei}
\affiliation{Space Sciences Laboratory, University of California, Berkeley, CA 94720, USA}

\author[0000-0002-9954-4707]{Jia Huang}
\affiliation{Space Sciences Laboratory, University of California, Berkeley, CA 94720, USA}

\author[0009-0008-0922-3995]{Mahdi Najafi-Ziyazi}
\affiliation{Centre for Mathematical Plasma-Astrophysics, Department of Mathematics, KU Leuven, Celestijnenlaan 200B, 3001 Leuven, Belgium}

\author{Hao Wu}
\affiliation{Centre for Mathematical Plasma-Astrophysics, Department of Mathematics, KU Leuven, Celestijnenlaan 200B, 3001 Leuven, Belgium}

\author{Rui Zhuo}
\affiliation{Centre for Mathematical Plasma-Astrophysics, Department of Mathematics, KU Leuven, Celestijnenlaan 200B, 3001 Leuven, Belgium}

\author{José M. L. Murteira}
\affiliation{Centre for Mathematical Plasma-Astrophysics, Department of Mathematics, KU Leuven, Celestijnenlaan 200B, 3001 Leuven, Belgium}

\author{Ketevan Arabuli}
\affiliation{Centre for Mathematical Plasma-Astrophysics, Department of Mathematics, KU Leuven, Celestijnenlaan 200B, 3001 Leuven, Belgium}

\author[0000-0003-3364-9183]{Brigitte Schmieder}
\affiliation{Centre for Mathematical Plasma-Astrophysics, Department of Mathematics, KU Leuven, Celestijnenlaan 200B, 3001 Leuven, Belgium}
\affiliation{Observatoire de Paris, LIRA, UMR8254 (CNRS), F-92195 Meudon Principal Cedex, France}
\affiliation{LUNEX EMMESI COSPAR-PEX Eurospacehub, Kapteyn straat 1, Noordwijk 2201 BB, Netherlands}

\author{Jasmina M. Magdalenić Zhukov}
\affiliation{Centre for Mathematical Plasma-Astrophysics, Department of Mathematics, KU Leuven, Celestijnenlaan 200B, 3001 Leuven, Belgium}
\affiliation{Solar-Terrestrial Centre of Excellence—SIDC, Royal Observatory of Belgium, 1180 Brussels, Belgium}

\begin{abstract}
In this paper, we simulate the magnetic flux evolution at different heliocentric distances during two solar-maximum Carrington rotations (CRs) using the time-evolving coronal magnetohydrodynamic (MHD) model COCONUT to investigate the ``open flux problem".
The simulated open magnetic flux (OMF) near the solar surface is comparable to that derived from \textit{in situ} observations by PSP and WIND satellites, and is about 5 times larger than that derived from  SDO coronal hole (CH) observations, and the variation in the simulated radial solar wind speed is consistent with the evolution of the OMF evaluated around the corresponding solar disk center. We find that the OMF is reduced by up to $45\%$ from 1.01~$R_s$ to 0.1~AU and increases with a higher-resolution mesh. The OMF decreases mainly within 3~$R_s$, where the closed magnetic flux drops more rapidly, from about $60\%$ of the total magnetic flux at 1.01~$R_s$ to about $4\%$ at 3~$R_s$. Moderate adjustment of the heating source term can effectively regulate the simulated OMF. Preprocessing the photospheric magnetograms with a potential field solver that removes many high-order spherical harmonic components reduces the OMF in the low corona, while having little impact beyond 3~$R_s$. Additionally, the ratio of the maximum to the minimum OMF can reach 1.4 during a single solar maximum CR. 
These findings highlight the necessity of considering higher grid resolution, more realistic heating mechanisms, and the time-evolving regime of coronal MHD modeling when further addressing the ``open flux problem". 
\end{abstract}

\section{Introduction}\label{sec:intro}
Most of the magnetic flux threading the photosphere returns to the solar surface, forming closed magnetic field lines in the corona, whereas the open magnetic field (OMF) extends into interplanetary space, linking the Sun and Earth and forming the interplanetary magnetic field. The OMF drives geomagnetic activity that affects daily life, guides the propagation of solar energetic particles, and helps shield the solar system from galactic cosmic rays \cite[e.g.][]{Schrijver2003,Erdos_2012,Linker_2017,Frost2022}.
Additionally, the relationship between long-term variations in solar OMF, solar activity, and potentially climate change remains an active area of investigation
\citep{Erdos_2012}. It is important to understand the mechanisms that drive the evolution of open magnetic flux.

It is widely accepted that coronal holes (CHs), characterized by rapidly expanding open field lines and appearing dark in extreme ultraviolet and X-ray images \citep{Linker1999JGR,CRANMER20023,FengMa2015,Feng_2017,FengandLiu2019}, are the primary sources of OMF. However, it was noted that the OMF derived from observation-based CH regions in synoptic line-of-sight photospheric magnetograms is smaller than that derived from interplanetary \textit{in situ} observations. Moreover, the OMF extrapolated numerically from the observed magnetograms into interplanetary space is typically underestimated by a factor of two or more. This discrepancy is widely known as the ``open flux problem", a well‐known challenge in coronal modeling \citep{Linker_2017}. 
It has been proposed that magnetic maps derived from observations may underestimate the total magnetic flux \citep[e.g.][]{Sinjan2024} or that a substantial fraction of the open magnetic flux originates from regions that do not appear dark in emission \citep{Linker_2017,Linker_2021}. 

The OMF can be evaluated by directly measuring the magnetic flux density using magnetometers onboard space probes. 
\textit{Ulysses} \citep{Wenzel1992,Smith1995} observations indicated that the magnetic flux density remains nearly uniform with the heliographic latitude at a given heliocentric distance \citep[except near the heliospheric current sheet;][]{Balogh1995,Smith1995,Lockwood2004}. 
\citet{Owens2008} further validated this finding by estimates of total heliospheric magnetic flux obtained from widely separated spacecraft, 
and revealed a tendency for the estimated flux to increase with heliocentric distance, which may be caused by inversions of heliospheric magnetic field lines \citep{Erdos_2012,Erdos_2014,Lockwood2009,Owens2017JGR,Frost2022}. This variation mainly arises from magnetic field fluctuations around the Parker spiral structure \citep{Erdos_2012}, due to $B_r$ decreasing with the heliospheric distance faster than the fluctuation amplitude \citep{Smith2011}. It was also reported that the warped current sheet, along with transient events such as co-rotating interaction regions and CMEs, can induce tilts in the heliospheric magnetic field, thereby contributing to the measured excess magnetic flux \citep{Lockwood2002}.

By adopting the supra-thermal electron beam method to distinguish between different heliospheric magnetic field topologies, it was found that magnetic field line inversions account for approximately $20\%$ of the measured magnetic flux in ACE data \citep{McComas1998,Smith1998,king2005JGR} collected during 1998-2011 \citep{Owens2017JGR}. Kinematic corrections based on the observed solar wind velocity structure have been proposed to mitigate the influence of large-scale spatial velocity gradients on OMF estimates \citep{Lockwood2009}. 
The \textit{in situ} electron and magnetic field data is also used to determine the global topology of the heliospheric magnetic field and identify field line inversions. With the development of these technologies, it has been demonstrated that the best heliospheric estimate of OMF remains, on average, a factor of 1.6 higher than the values extrapolated from photospheric observations \citep{Frost2022}. 

By measuring the magnetic field observed by the FIELDS instrument \citep{Bale2016} onboard the Parker Solar Probe \citep[PSP;][]{Fox2016} along the Parker spiral, and then projecting it radially to correct for excess flux caused by field line inversions \citep{Erdos_2012,Erdos_2014}, it was found that the total unsigned \textit{in situ} magnetic flux remains nearly constant between 0.13~AU and 0.8~AU \citep{Badman2021}. 
From MHD simulations, \cite{Riley2021} showed that the underestimated magnetic flux persists even down to 26.9~$R_s$. 
Additionally, it was reported that the magnetic field may continue to vary along the latitude out to at least $10~R_s$ \citep{Reville_2017}, highlighting the necessity of modeling the evolution of open flux down to the trans-Alfv{\'e}nic region in order to assess how magnetic flux evolves with heliospheric distance. 

Using CH detections, the OMF can be estimated by integrating the magnetic flux within the CH regions identified on photospheric magnetograms. 
However, OMF also presents outside CHs and transports across the solar surface via interchange reconnection with closed magnetic loops \citep{Fisk2005,Iijima_2025}. The complex structure and dynamic interactions between closed and open magnetic fields near the CH boundary introduce uncertainties in CH detection.
Even when advanced CH detection techniques are applied to MHD simulation results, there remains a tendency to overestimate the total open-flux area, but miss the small CHs associated with active regions. Implemented in the simulation results, it underestimated the total simulated open magnetic flux by $\sim 30$-$40 \%$ \citep{Linker_2021}, still insufficient to account for the observed missing flux.

Magnetograph measurements may also lead to missed OMF. 
The Zeeman-Doppler imaging (ZDI) magnetograms of stellar surface magnetic fields suffer from missing information in regions obscured by dark starspots or hidden from view due to the rotation axis' inclination \citep{Jardine2010,Johnstone2010}. The magnetic field that ZDI fails to recover in these regions, or misses due to limited spatial resolution, may lead to an underestimation of the OMF. By analyzing simulated solar photospheric magnetic fields, \cite{Milic2024} and \cite{Sinjan2024} showed that reduced telescope resolution leads to a lower estimated mean magnetic flux density and recommended using high-resolution observations for magnetic field extrapolations to mitigate the effect of underestimated unipolar magnetic flux. Additionally, \citet{Wang_2022OF} showed that by applying saturation corrections \cite[e.g.,][]{Ulrich2019} to MWO and WSO photospheric magnetograms and then extrapolating the magnetic field from them with the PFSS model, the derived OMF can be consistent with the observed interplanetary magnetic flux. \citet{wang2025sipifvmobservationbasedmagnetohydrodynamicmodel} showed that inserting an observation-derived flux-rope field into the inner-boundary magnetic field leads to a larger open-field region that is more consistent with CH observations.

There are also simulation research works related to the ``open flux problem". Potential field source surface (PFSS) simulations indicated that the total non-dipole flux component is at least an order of magnitude larger than the dipole flux \citep{Yoshida_2023}.
\citet{Riley2019} added additional polar flux to photospheric magnetograms to (partially) resolve the ``open flux problem" in PFSS and MHD simulations during solar minimum. 
\cite{Wallace2019} showed that the total OMF derived from the observed CHs agrees well with the semi-empirical Wang-Sheeley-Arge (WSA) simulation results \citep{Wang1990ApJ,ARGE20041295}, and both deviated from the \textit{in situ} observations, especially near solar maximum.
\citet{Arge_2024} showed that adjusting the boundary by about one supergranular width, a key scale in open–closed flux dynamics \citep{Aslanyan_2022}, is sufficient to align the modeled OMF in the WSA simulations with \textit{in situ} observations.
\citet{Caplan_2021} found that increasing the grid resolution in PF simulations reduces open-field areas, increases the unsigned magnetic flux, and hardly impacts the net OMF, and showed that the choice of source surface radius strongly influences the calculated OMF. \citet{wang2025COCONUTMayEvent} indicated that increasing the tangential
grid resolution by \(4 \times 4\) times (from
\(\sim3.6^{\circ}\) to \(\sim0.9^{\circ}\)) in coronal MHD simulations led to an increase of
$\sim40\%$ in the average magnetic field strength. Besides, \citet{Asvestari_2024} showed that the MHD model produces significantly more OMF than the PF models.

Moreover, \citet{Yeates2010} and \citet{Mackay_2022} inserted magnetic bipoles, determined from observed synoptic magnetograms, into their coronal simulations to provide a continuously evolving boundary condition for the non-potential magnetofrictional (MF) model. It is observed that surface motions drive the twisted magnetic flux ropes to erupt and induce fluctuations in the OMF. 
High-cadence bipole data derived from hourly updated advective flux transport (AFT) magnetograms \citep{Upton2014a} produced more OMF than the 27-day Carrington-rotation maps, and non-potential simulations yield substantially higher OMF than PFSS models \citep{Mackay_2022}. Given that MF models cannot capture the dynamic features of the plasma, time-evolving MHD coronal simulations are required to investigate magnetic flux evolution in a more self-consistent manner. Consequently, coronal MHD models that can simultaneously achieve high computational efficiency, strong numerical stability, and high spatial and temporal resolution are required to perform such simulations. 

In general, MHD models of the solar corona and wind can be classified into two categories: quasi-steady models and time-evolving models. Models constrained by a single static magnetogram \citep[e.g.][]{Perri2018SimulationsOS,Perri_2022,Perri_2023,MIKIC2018NatA,WANG201967,Feng2005,Feng_2010,Feng_2021,Parenti_2022,Kuzma_2023,Linan_2023,Liu_2023,Wang_2022,Wang2022_CJG,wang2025sipifvmobservationbasedmagnetohydrodynamicmodel,WangSIPtheoriticalCME,brchnelova2023role} are referred to as quasi-steady-state models, whereas those driven by a sequence of time-evolving magnetograms \citep[e.g.][]{Yang2012,Hayashi_2021,Hoeksema2020,Feng_2023,Lionello_2023,Mason_2023,wang2025COCONUTMayEvent,Wang2025_FirsttimeevolvingCOCONUT,wang2025sipifvmtimeevolvingcoronalmodel} are defined as time-evolving models. The former assume a (quasi-)steady corona over one CR, whereas the latter are time-accurate and driven by continuously updated magnetograms, thereby enabling a more realistic representation of the evolving solar coronal and wind structures. 
A more detailed description of these two coronal modeling regimes is available in \cite{wang2025COCONUTMayEvent,Wang2025_FirsttimeevolvingCOCONUT,wang2025sipifvmtimeevolvingcoronalmodel}.

Currently, implicit temporal integration methods, which allow selection of time steps larger than the explicit method,
have significantly improved the computational efficiency of quasi-steady-state coronal MHD models \citep{Perri_2022,Perri_2023,WANG201967,Feng_2021,Wang_2022,Wang2022_CJG}, achieving tens- to hundreds-fold speedups compared with explicit models. Moreover, the time-evolving coronal MHD models COolfluid COroNal UnsTructured \citep[COCONUT;][]{wang2025COCONUTMayEvent,Wang2025_FirsttimeevolvingCOCONUT,Wang_COCONUT_DecE} and Solar Interplanetary Phenomena Implicit Finite Volume Method \citep[SIP-IFVM;][]{wang2025sipifvmtimeevolvingcoronalmodel}, both adopting implicit methods, realize faster-than-real-time simulations by only about twenty CPUs, with grid resolutions of 1.5 million (M) and 1 M cells and time steps of 10 and 3-4 minutes, respectively. 
Whereas the commonly used explicit or semi-implicit time-evolving coronal MHD models \citep{Hoeksema2020,Downs2025PSI} typically require thousands of CPUs to achieve comparable performance.
Moreover, the extended magnetic field decomposition method \citep{wang2025sipifvmtimeevolvingcoronalmodel} and the decomposed energy strategy
\citep{Wang_COCONUT_DecE} enable SIP-IFVM and COCONUT to perform well in time-evolving MHD coronal simulations involving low-$\beta$ (thermal-to-magnetic pressure ratio) issues \citep{Feng_2021, Wang_2022,wang2025sipifvmtimeevolvingcoronalmodel,Wang_COCONUT_DecE}.

In this paper, we use the time-evolving COCONUT, a novel implicit coronal model built on the Computational Object-Oriented Libraries for Fluid Dynamics (COOLFluiD) framework \citep{kimpe2,lani1,lani13}\footnote{\url{https://github.com/andrealani/COOLFluiD/wiki}}, to simulate the evolution of magnetic flux during two solar-maximum CRs, as a preparation for a comprehensive whole-solar-cycle MHD simulation of OMF evolution in the future. 
Inherent from high efficiency, COCONUT \citep{wang2025COCONUTMayEvent,Wang2025_FirsttimeevolvingCOCONUT} has finished a continuously evolving coronal simulation dataset that covers three years around the solar minimum to train the machine-learning neural network \citep{Li_2025}.
Given that the numerical stability of COCONUT has been significantly improved by the decomposed-energy strategy \citep{Wang_COCONUT_DecE}, we further utilize COCONUT to calculate magnetic evolutions around solar-maximum CRs to investigate the ``open flux problem".

Based on the above considerations, the paper is organized as follows. In Section~\ref{Numericalsetup}, we describe the numerical setup of COCONUT used in this study.
In Section \ref{sec:Numerical Results}, we present the evolution of the magnetic flux and the distributions of open-field regions during solar maximum CRs 2282 and 2283 simulated by COCONUT. Simulation results with different model setups are compared to assess uncertainties in coronal MHD modeling. Comparisons with interplanetary \textit{in situ} and Extreme Ultraviolet (EUV) CH observations are also provided. In the Appendix, we present the plasma evolution and the corresponding results used to validate our conclusions.
In Section~\ref{sec:Conclusion}, we summarize the simulation results and provide concluding remarks.

\section{Numerical setup in COCONUT}\label{Numericalsetup}
In this paper, we use the coronal MHD model COCONUT, which employs the decomposed energy strategy and the HLL Riemann solver with an additional dissipation term added to the energy equation to improve numerical stability in addressing low-$\beta$ issues \citep{Wang_COCONUT_DecE}, to simulate coronal evolutions during two solar maximum CR. The impact of several numerical modifications introduced in \cite{wang2025COCONUTMayEvent,Wang2025_FirsttimeevolvingCOCONUT}, originally developed to improve numerical stability, is also evaluated using this version of COCONUT.

\subsection{The governing equations and grid system}\label{Governingequations} 
A spherical-shell computational domain extending from 1.01 to $\sim25\,R_s$, discretized into a sixth-level subdivided geodesic mesh \citep{Brchnelova2022} containing about 1.5 million (M) truncated pentagonal‑pyramid cells, with a tangential angular resolution of approximately $1.8^\circ$ \citep{wang2025COCONUTMayEvent,Wang2025_FirsttimeevolvingCOCONUT}, is adopted for the simulations. The governing equations are the same as those in \citep{Wang_COCONUT_DecE} and can be described in the following form:
$$
\frac{\partial \mathbf{U}}{\partial t}+\nabla \cdot \mathbf{F}\left(\mathbf{U}\right)=\mathbf{S}\left(\mathbf{U},\nabla \mathbf{U}\right) \,,
$$
where $t$ and $\mathbf{U}$ refer to the time and vector of conservative variables, $\nabla \mathbf{U}$ means the spatial derivative of $\mathbf{U}$, and $\mathbf{\mathbf{F}\left(\mathbf{U}\right)}$ is the inviscid flux vector, $\mathbf{S}\left(\mathbf{U},\nabla \mathbf{U}\right)=\mathbf{S}_{\rm gra}+\mathbf{S}_{\rm heat}+\mathbf{S}_{\rm{DECOMP}}$ denotes the vector of the source terms corresponding to the gravitational force, the heating source terms, and the source term derived from the decomposed energy equation described in \cite{Wang_COCONUT_DecE}. We calculate $\mathbf{S}_{\rm gra}$ and $\mathbf{S}_{\rm DECOMP}$ in the same way as before, but vary the calculation of $\mathbf{S}_{\rm heat}$ to evaluate its impact on global coronal MHD modeling. The term $\mathbf{S}_{\rm heat}=-\nabla \cdot \mathbf{q}+Q_{rad}+Q_{H}$ consists of optically thin radiative loss $Q_{rad}$, empirically defined coronal heating $Q_{H}$, and Spitzer or collisionless thermal conduction term $-\nabla \cdot \mathbf{q}$. 

As usual, $Q_{rad}$ is calculated the same as in \cite{Wang2025_FirsttimeevolvingCOCONUT,wang2025COCONUTMayEvent}. 
The thermal conduction term is computed following the approach described in \cite{Baratashvili2024} and \cite{Wang2025_FirsttimeevolvingCOCONUT}.
As usual, $Q_{H}$ can be calculated by the following formulation \citep{Mok_2005,Downs2010,Baratashvili2024,WangSIPtheoriticalCME,wang2025COCONUTMayEvent}:
\begin{equation}\label{Coronalheating}
Q_{H}=H_0 \cdot \left|\mathbf{B}\right| \cdot e^{-\frac{r-R_s}{\lambda}},~\text{with}~H_0 = 4 \cdot 10^{-2}\,{\rm J\,m^{-3}\,s^{-1}\,T^{-1}}\,,
\end{equation}
where $\mathbf{B}$ is the magnetic field vector, and $\lambda = 0.7\,R_s$. To evaluate the uncertainty associated with the empirically defined $Q_{H}$, we also tested the following slightly modified form for comparison:
\begin{equation}\label{Coronalheatingmodified}
Q_{H}=H_0 \cdot \left|\mathbf{B}\right| \cdot \frac{r}{R_s} \cdot e^{-\frac{r-R_s}{\lambda}},~\text{with}~H_0 = 2 \cdot 10^{-2}\,{\rm J\,m^{-3}\,s^{-1}\,T^{-1}}\,.
\end{equation} 
Inspired by the empirically defined volumetric heating source terms \citep[e.g.,][]{Nakamizoetal2009,Feng_2010,Feng_2021,Wang_2022}, which introduce a dimensionless heliocentric distance parameter into the exponentially decaying term, we further scale Eq.~(\ref{Coronalheating}) by $\frac{r}{R_s}$ to obtain Eq.~(\ref{Coronalheatingmodified}). 
Because $\frac{r}{R_s} \ge 1$, multiplying the heating term by this factor increases the prescribed heating. To avoid introducing an unrealistically large energy input, we reduce $H_0$ in Eq.~(\ref{Coronalheatingmodified}) so that the overall heating remains comparable to the original formulation.
This comparison is not intended to identify a superior empirical heating source term, but rather to assess how radial variations in heating prescriptions alter the simulated coronal structures.

\subsection{Inner boundary setups}\label{BCinner}
A series of hourly-updated photospheric magnetograms is used to drive coronal evolution. These simulations are conducted in a heliocentric inertial coordinate system \citep{Burlaga1984MHDPI,FRANZ2002217} with the Earth permanently positioned at $\phi=60^{\circ}$. During the time-evolving coronal simulations, the inner-boundary radial magnetic field at each physical time step is interpolated from four adjacent observation-based magnetograms using cubic Hermite interpolation \citep{Wang2025_FirsttimeevolvingCOCONUT}. 

Since the chromosphere and transition region are not included in our model, the inner-boundary radial magnetic field $B_{r,s}$ is defined by employing a PF solver with the 0th- and high-order spherical harmonic components removed to extrapolate the photospheric magnetograms to the base of the low corona. 
To evaluate the impact of this preprocessing, which smooths small-scale magnetic structures, we use two magnetogram datasets to drive coronal evolution. 
\begin{enumerate}
      \item  We use a 10th-order spherical-harmonic PF solver, in which the spherical-harmonic expansion of the scalar potential $\mathcal{U}$ is given by
\begin{equation}\label{originalspherical}
\mathcal{U} \approx \sum\limits_{l=1}^{10} \sum\limits_{m=-l}^{l} I_l^m Y_l^m(\mu, \phi) \, .
\end{equation}
      \item  We employ a 50th-order spherical-harmonic PF solver with a filter 
\citep{MCCLARREN20105597, wang2025COCONUTMayEvent, Murteira2025}, expressed as
\begin{equation}\label{filteredspherical}
\mathcal{U}_{\rm filtered} \approx \sum\limits_{l=1}^{50} \sum\limits_{m=-l}^{l} \frac{I_l^m Y_l^m(\mu, \phi)}{1 + \xi\, l^{2}(l+1)^{2}} \,, \text{with}~\xi = 3 \times 10^{-6}.
\end{equation} 
   \end{enumerate}
In Eqs.~(\ref{originalspherical}) and (\ref{filteredspherical}), $Y_l^m$ and $I_l^m$ denote the spherical harmonic function and corresponding spectral coefficient of degree $l$ and order $m$, $\mu \equiv \cos\theta$, $\theta \in [0, \pi]$ and $\phi \in [0, 2\pi)$ present the colatitude and longitude. 

The inner boundary velocity and tangential magnetic field, denoted by $\mathbf{v}_s$, $B_{\theta,s}$ and $B_{\phi,s}$, are defined the same as in \cite{wang2025COCONUTMayEvent,Wang2025_FirsttimeevolvingCOCONUT}, the pressure is set to $p_s=0.01\;\rm Pa$, and the inner boundary plasma density $\rho_s$ is defined as a piecewise polynomial in the magnetic field strength $|\mathbf{B}_s|=\sqrt{B_{r,s}^2+B_{\theta,s}^2+B_{\phi,s}^2}
$~:
\begin{equation}\label{inhomogeniousdensity}
\rho_s(|\mathbf{B}_s|)=\left\{\begin{array}{c}
\frac{\rho_{\rm ref1}-\rho_{\rm ref}}{B_{\rm ref1}}\cdot |\mathbf{B}_s|+\rho_{\rm ref},  \text { if } |\mathbf{B}_s| \leq B_{\rm ref1} \\
\\
\frac{\left(\rho_{\rm ref2}-\rho_{\rm ref1}\right) \cdot \left(|\mathbf{B}_s|-B_{\rm ref1}\right)}{B_{\rm ref2}-B_{\rm ref1}}+\rho_{\rm ref1},  \text { if } B_{\rm ref1} \le |\mathbf{B}_s| < B_{\rm ref2} \\
\\
\frac{\left(\rho_{\rm ref3}-\rho_{\rm ref2}\right) \cdot \left(|\mathbf{B}_s|-B_{\rm ref2}\right)}{B_{\rm ref3}-B_{\rm ref2}}+\rho_{\rm ref2},  \text { if } B_{\rm ref2} \le |\mathbf{B}_s| 
\end{array}\right.\,,
\end{equation}
where $\rho_{\rm ref}=3.34\times 10^{-13}\;\rm kg~m^{-3}$ is the inner boundary density adopted in \cite{wang2025COCONUTMayEvent}. $\rho_{\rm ref1}=3.56\times 10^{-13}\;\rm kg~m^{-3}$, $\rho_{\rm ref2}=16.94\times 10^{-13}\;\rm kg~m^{-3}$, and $\rho_{\rm ref3}=135.96\times 10^{-13}\;\rm kg~m^{-3}$ denote the plasma densities at 6 Mm above the photosphere derived from three surface-to-corona simulations performed with the \textit{Bifrost} radiative MHD code \citep{Gudiksen2011}. These aim at spanning different typical configurations, from weakly magnetized quiet-Sun conditions to network-like emergence episodes \citep{NorazChromosphereQuietSun2026}, each configured with photospheric magnetic field of $B_{\rm ref1}=21$ G, $B_{\rm ref2}=61$ G, and $B_{\rm ref3}=89$ G, respectively.

Since the numerical stability of COCONUT in addressing low-$\beta$ issues has been substantially improved by \cite{Wang_COCONUT_DecE}, we remove the pressure positivity-preserving (PP) treatment \citep{wang2025COCONUTMayEvent} to minimize the influence of algorithmic uncertainties. Meanwhile, we retain the PP measure on density \citep{wang2025COCONUTMayEvent} to prevent potential code failures in the challenging time-evolving solar-maximum CR simulations. 

\section{Numerical results}\label{sec:Numerical Results}
In this section, we evaluate the evolution of magnetic flux during CRs 2282 and 2283 at different heliocentric distances. In Appendix \ref{sec:EvolutionofCoronalStructure}, we present the simulated coronal evolution in three cases and compare the results with observations to validate the model and assess the impact of numerical uncertainties on the simulated coronal structures. In Appendix \ref{Filterornot}, we illustrate the impact of filtered and unfiltered PF solver preprocessing of initial photospheric magnetograms on the simulated open field distribution.

About 1300 hourly-updated GONG-zqs photospheric magnetograms\footnote{\url{https://gong.nso.edu/data/magmap/QR/zqs/}} from CRs 2282 and 2283, spanning from  22:14 on March 12, 2024, to 11:14 on May 6, 2024, are used to drive the coronal evolutions. To achieve an appropriate balance between accuracy and computational efficiency in the time-evolving coronal simulations, we use a time step of 5 min. Three time-evolving simulations using different heating source terms and photospheric magnetogram preprocessing methods as listed bellow are compared to evaluate numerical uncertainties. 
 \begin{enumerate}
     \item Case~1: Eq.~(\ref{Coronalheating}) for heating source term and Eq.~(\ref{originalspherical}) for magnetogram preprocessing.
     \item Case~2: Eq.~(\ref{Coronalheating}) for heating source term and Eq.~(\ref{filteredspherical}) for magnetogram preprocessing.
     \item Case~3: Eq.~(\ref{Coronalheatingmodified}) for heating source term and Eq.~(\ref{filteredspherical}) for magnetogram preprocessing.
\end{enumerate}

For case~1, the maximum magnetic field strength and the minimum plasma-$\beta$ in the low corona around active regions are typically about 17~G and $9\times10^{-3}$, respectively, and can occasionally reach values as high as 27~G and as low as $3\times10^{-3}$.
For Cases~2 and~3, the maximum magnetic field strength and the minimum plasma-$\beta$ are typically around 60~G and $6\times10^{-4}$, respectively, and can occasionally reach values as high as 99~G and as low as $3\times10^{-4}$. 
All simulations are performed using 720 parallel processes on the WICE cluster, which is part of the Tier-2 supercomputer at the Vlaams Supercomputer Centrum\footnote{\url{https://www.vscentrum.be/}}. Under this configuration, the simulations run approximately 35-40 times faster than real-time evolutions.

Magnetic field lines are traced from $150 \times 300$ uniformly distributed seed points on a spherical surface at 10~$R_s$ in both the sunward and the anti-sunward directions, and a field line is classified as open at $r = r_{\rm se}$ if it reaches both $r_{\rm se}$ and 15~$R_s$. 
After determining the footpoints of the open magnetic field lines at $r = r_{\rm se}$, we compute the distance between each sample point and the set of derived footpoints. In this work, $150 \times 300$ uniformly distributed sample points on the spherical surface at $r = r_{\rm se}$ are adopted, and a sample point is classified as belonging to an open-field region if the minimum distance is less than $0.01\,\pi\cdot R_{\rm se}$.  The unsigned OMF at $R_{\rm se}$ is calculated as $$\Phi_{{\rm open~at}~R_{\rm Se}}=\sum\limits_{i=1}^{N^{{\rm open~at~}R_{\rm Se}}}|B^{{\rm open~at}~R_{\rm Se}}_{r,i}|\cdot{\rm S}^{{\rm open~at~}R_{\rm Se}}_i$$ 
where $B^{{\rm open~at}~R_{\rm Se}}_{r,i}$ and ${\rm S}^{{\rm open~at~}R_{\rm Se}}_i$ denote the radial magnetic field at the $i$th sample point within the open-field region and the discretized area associated with that point, respectively, and $N^{{\rm open~at~}R_{\rm Se}}$ denotes the number of the sample points belonging to open-field regions at $R_{\rm Se}$.

\begin{figure*}[htpb]
\begin{center}
  \vspace*{0.01\textwidth}
    \includegraphics[width=0.9\linewidth,trim=1 1 1 1, clip]{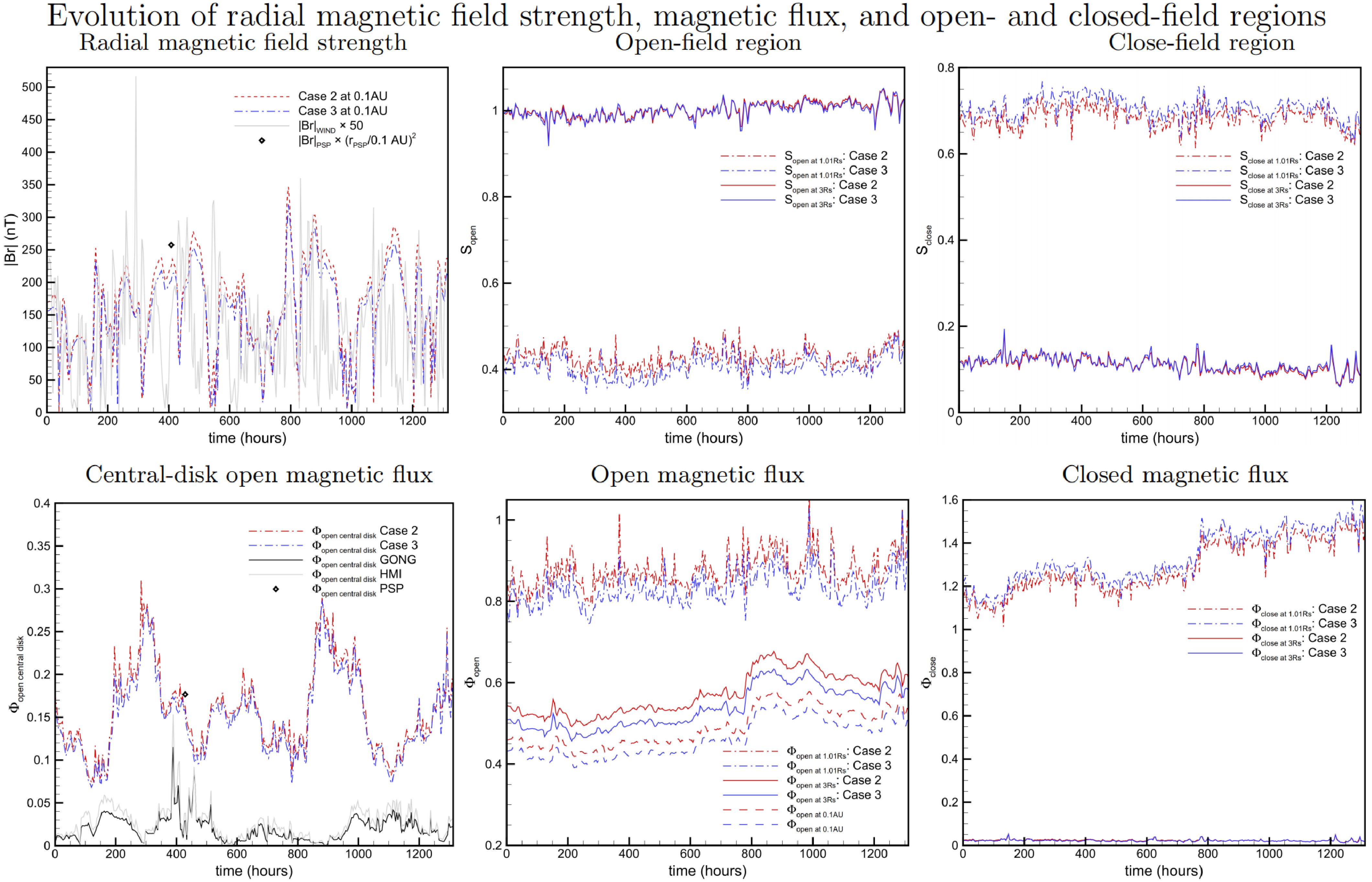}
\end{center}
\caption{Timing diagrams of the radial magnetic field strength from the WIND observations scaled by $0.5\cdot \left(\frac{1~\rm AU}{0.1~\rm AU}\right)^2$ and measured by a virtual satellite located at 21.5~$R_s$ with the same latitude as Earth but lags by $60^\circ$ in longitude (nT; top left), of the areas of the simulated open- and closed-field regions ($\rm S_{open}$ and $\rm S_{close}$; top middle and right), of the unsigned open and closed magnetic flux ($\Phi_{\rm open}$ and $\Phi_{\rm close}$; bottom middle and right) at 1.01~$R_s$, 3~$R_s$ and 0.1 AU, and of the unsigned OMF at 1.01~$R_s$ covering a region of $45^{\circ}$ from the solar disk center in longitude and latitude as viewed from Earth, together with the unsigned flux derived from CH regions identified in SDO observations within the corresponding disk center region ($\Phi_{\rm open~central~disk}$; bottom right).
The magnetic fluxes are normalized using the WIND \textit{in situ} observations,
$\Phi_{\rm 1AU}^{\rm ave}
=
\left| B_{r,\rm 1AU} \right|^{\rm ave}
\,\cdot\,4\pi\,(1~\mathrm{AU})^{2}=7.67 \times 10^{14}~{\rm Wb}$,
where $\left| B_{r,\rm 1AU} \right|^{\rm ave}$ denotes the average unsigned radial interplanetary magnetic field strength along the Sun-Earth direction derived from about 1310 hourly averaged WIND observations during the simulated period. 
The areas of the open-field regions at $r = r_{\rm Se}$ are normalized by
${\rm S}_{\rm open~at~3R_s}^{\rm ave}\cdot\left(r_{\rm Se}/3~R_s\right)^2$,
where ${\rm S}_{{\rm open~at}~3~R_s}^{\rm ave} = 4.92\times10^{19}~\mathrm{m}^2$ denotes the average open-field area at 3~$R_s$ derived from about 330 simulation results in Case~3, with a cadence of 4 hours.
The red, and blue curves correspond to results obtained from Cases~2, and~3, respectively. The diamond symbol in left panels denotes the results obtained from PSP observations, and the black and gray curves in the bottom‑left panel represent the observed OMF derived from the GONG and HMI magnetograms, respectively, within CHs identified from SDO EUV observations.
}\label{Fluxevolution}
\end{figure*}

\begin{table*}
\centering
\caption{ Averaged normalized magnetic flux variables over the simulated period}
\label{averagedopenflux}
\begin{tabular}{lllll}
\hline\noalign{\smallskip}
Case & $\overline{\Phi}_{\rm total}$ at 1.01 \& 3~$R_s$~~&~~ $\overline{\Phi}_{\rm open}$ at 1.01,~3~\& 21.5~$R_s$~~& $\overline{\rm S}_{\rm open}$ at 1.01~\& 3~$R_s$~~& $\overline{\rm S}_{\rm open~central~disk}$~\& $\overline{\Phi}_{\rm open~central~disk}$ at 1.01~$R_s$\\
\noalign{\smallskip}\hline\noalign{\smallskip}
1 & 1.910~~\&~~0.596 &~~~~0.771,~~0.577~~\&~~0.494 &~0.382~~\&~~1.012&~~~~0.048 ~~\&~~0.130\\
2  & 2.165~~\&~~0.597 &~~~~0.875,~~0.573~~\&~~0.491 &~0.427~~\&~~1.002&~~~~0.046 ~~\&~~0.158\\
3  & 2.165~~\&~~0.560 &~~~~0.833,~~0.538~~\&~~0.461 &~0.405~~\&~~1.000&~~~~0.043 ~~\&~~0.149\\
\noalign{\smallskip}\hline
\end{tabular}
\end{table*}

In Figure~\ref{Fluxevolution}, the top-left panel shows that the fluctuation frequency of the simulated radial magnetic field strength is basically consistent with the observations, whereas its magnitude is approximately only half of the observed value at 1~AU scaled by $\left(0.1~\mathrm{AU}/1~\mathrm{AU}\right)^2$. 
This is consistent with the result at bottom-middle panel showing that the simulated OMF at 0.1~AU is also approximately half of the averaged interplanetary magnetic flux derived from the WIND observations \footnote{\url{https://cdaweb.gsfc.nasa.gov/index.html/}} \citep{king2005JGR}, assuming latitudinal invariance of the interplanetary magnetic field distribution. In addition, the simulated radial magnetic field strength at the 408th hour of the simulation, corresponding to 22:15 UT on March 29 when the latitude of PSP satellite aligns with that of the virtual satellite, is 227.0, 224.9, and 202.4 nT in Cases 1, 2, and 3, respectively. These values are broadly consistent with the PSP observation of 257.4 nT when scaled from 0.05 AU to 0.1 AU. 
From the bottom-middle panel of Figure~\ref{Fluxevolution} and Table \ref{averagedopenflux}, it is also noticed that the OMF derived from the simulations at 1.01~$R_s$ in Cases~2 and~3 amounts to approximately $85\%$ of the interplanetary magnetic flux $\Phi_{\rm 1AU}^{\rm ave}$, and about $77\%$ in Case~1.
Here, $\Phi_{\rm 1AU}^{\rm ave} = 7.67 \times 10^{14}$~Wb is derived from hourly averaged WIND \textit{in situ} observations during CRs 2282 and 2283.
Considering that hourly averaged interplanetary magnetic field measurements tend to overestimate the open magnetic flux due to the commonly occurring inversions of the interplanetary magnetic field, and that the ratio between the OMF inferred from hourly averaged \textit{in situ} observations and that derived from daily averaged data which mitigate the effects of magnetic field inversions is $137\%$ (the corresponding ratio reported by \citet{Linker_2021} is approximately $130\%$), we conclude that the simulated OMF evaluated in the low corona is generally consistent with the interplanetary observations. 

The bottom-left panel of Figure~\ref{Fluxevolution} shows that the temporal variation of the central disk OMF, evaluated in a region of $45^{\circ}$ from the solar disk center in longitude and latitude as viewed from Earth, is consistent with the radial velocity at 0.1 AU shown in Figure~\ref{1DTimingDiagram}, reflecting that the high-speed solar wind is dominated by OMF. Together with Table~\ref{averagedopenflux}, we find that the ratio of central-disk to full-spherical open-field regions is significantly less than the corresponding ratio of OMF. This is because the magnetic field strength in low-latitude open-field regions is higher than that in high-latitude regions. Although the temporal evolution of the central-disk OMF exhibits a periodicity of one CR, significant variations in the shape and amplitude of the periodically appeared peaks and troughs reveal substantial temporal fluctuations. This underscores the importance of multi-satellite observations for constructing more realistic, time-synchronized magnetograms, thereby improving coronal modeling during solar-maximum CRs. 

We also compare the OMF obtained from MHD simulations with that derived from EUV CH observations in the bottom-left panel of Figure~\ref{Fluxevolution}. CH boundaries within the corresponding central-disk regions are extracted from SDO/AIA 193\,\AA\ observations using the Detection and Tracking Algorithm for Coronal Holes (DETACH; \citealt{liuDETACHDetectionTracking2026}). These delineated regions are then mapped onto photospheric magnetograms to calculate the OMF for individual CHs, which are subsequently summed to obtain the total unsigned OMF across the entire central-disk region. The calculations are independently carried out using magnetic field observations from both the GONG and HMI instruments, which possess differing spatial resolutions. This indicates that the OMFs derived from EUV CH observations are not largely sensitive to the spatial resolution of the underlying magnetic field data and are consistently lower than those obtained from MHD simulations. In particular, several prominent peaks in the temporal evolution of the central-disk OMF observed in the simulation results are absent in the CH-based observations. 
This divergence can be attributed to the inherent limitations of EUV-based CH identification methods. Some OMF do not originate from visually dark CHs \citep{Asvestari_2024, ngampoopunInvestigatingSolarWind2025}. For example, regions between active regions and CH boundaries, where continuous energy transport, driven by processes such as interchange magnetic reconnection, can lead to elevated plasma densities and temperatures \citep{heinemannOriginSuddenHeliospheric2024,Iijima_2025}. Consequently, these open-field regions exhibit significantly higher EUV brightness than typical CHs and are therefore excluded by the commonly used EUV segmentation techniques, leading to discrepancies with the simulation results. 

Furthermore, we validate our simulation results by comparing with the PSP \textit{in situ} observations. We derive the central-disk OMF at the $t$th hour of the simulation from PSP \textit{in situ} observations as follows.
$$\Phi^{\rm PSP,t}_{\rm open~central~disk}=\frac{\int_{\phi_{t}-0.25\pi}^{{\phi_{t}+0.25\pi}}\left| B^{\rm PSP}_{r,\phi} \right|\cdot\left(\frac{\rm 0.1~AU}{r}\right)^2\cdot \left|d\phi\right|} {\int_{\phi_{t}-0.25\pi}^{{\phi_{t}+0.25\pi}}\left|d\phi\right|}
\,\cdot\,\frac{\sqrt{2}}{2}\pi\,\cdot\,(0.1~\mathrm{AU})^{2}\,,$$
where $\phi_t$ denotes the longitude of PSP in the CR coordinate system at the $t$th hour of the simulation, $B^{\rm PSP}_{r,\phi}$ represents the radial magnetic field strength measured by PSP at longitude $\phi$ and heliocentric distance $r$, corresponding, for example, to the $t^{\prime}$th hour of the simulation, and $\left|d\phi\right|$ denotes half of the angular distance between the two PSP sampling longitudes closest to the $t^{\prime}$th hour. We note that around March 29, 2024, PSP is near perihelion and sweeps across longitudes from $99.5^\circ$ to $169.5^\circ$ in the CR coordinate system between 04:15 UT on March 30 and 00:00 UT on April 1. Coincidentally, Earth is located at a longitude of $124.5^\circ$ in the CR frame at 19:00 UT on March 30. Accordingly, we compare the central-disk OMF derived from PSP \textit{in situ} observations over 04:15 UT on March 30 and 00:00 UT on April 1 with that obtained from SDO CH observations at 19:00 UT on March 30, corresponding to the simulation result at the $428.8$th hour. The results in the bottom-left panel of Figure~\ref{Fluxevolution} demonstrate that our simulation results well capture the PSP \textit{in situ} observations.

In addition, the bottom-middle panel in Figure~\ref{Fluxevolution} shows that, in Cases~2 and~3, the OMF derived from the simulations decreases from $88\%$ and $83\%$ to $57\%$ and $54\%$ of the interplanetary magnetic flux $\Phi_{\rm 1AU}^{\rm ave}$ between 1.01~$R_s$ and 3~$R_s$, and is further reduced by about $7\%$ of $\Phi_{\rm 1AU}^{\rm ave}$ from 3~$R_s$ to 0.1~AU. Besides, the top-middle panel of Figure~\ref{Fluxevolution}, together with Table \ref{averagedopenflux}, reveals that the fraction of open-field regions at 3~$R_s$ is approximately 2.6, 2.3, and 2.5 times that at 1.01~$R_s$ for Cases~1,~2, and~3, respectively. The top-right and bottom-right panels of Figure~\ref{Fluxevolution} show that the normalized closed-field regions reduce from about 0.7 at 1.01~$R_s$ to 0.1 at 3~$R_s$, and the closed-field flux even reduced from 1.3 at 1.01~$R_s$ to less than 0.05 at 3~$R_s$. The ratio of open to total unsigned magnetic flux increases from about $40\%$, $40\%$, and $38\%$ at 1.01~$R_s$ to approximately $97\%$, $96\%$, and $96\%$ at 3~$R_s$ for Cases~1,~2, and~3, respectively. This indicates a much more rapid decrease in closed magnetic field structures than the open field in the low coronal region with increasing heliocentric distance. Furthermore, we conduct a simulation using a seventh-level grid mesh \citep{wang2025COCONUTMayEvent}, which contains $2\times2$ times as many discretized grid cells as the sixth-level mesh in tangential direction and remains the same in radial direction, and compare the simulated magnetic flux, as listed in Table \ref{openfluxLV6vsLV7}. Compared with the level-6 mesh, the level-7 mesh increases the total unsigned magnetic flux by $2.2\%$ and $10.6\%$ at $1.01~R_s$ and $3~R_s$, respectively, while increases the OMF by $12.6\%$, $12.1\%$, and $14.4\%$ at $1.01~R_s$, $3~R_s$, and 0.1 AU, respectively. A higher grid resolution increases both the area of open field regions and the magnitude of the OMF, and meanwhile reduces the closed magnetic flux in the low corona. 
 
\begin{table*}
\centering
\caption{ Normalized magnetic flux variables at the initial time for level-6 and level-7 meshes in Case~2 }
\label{openfluxLV6vsLV7}
\begin{tabular}{llll}
\hline\noalign{\smallskip}
Mesh & $\Phi_{\rm total}$ at 1.01 \& 3~$R_s$~~&~~ $\Phi_{\rm open}$ at 1.01,~3~\& 21.5~$R_s$~~& $\rm S_{\rm open}$ at 1.01~\& 3~$R_s$\\
\noalign{\smallskip}\hline\noalign{\smallskip}
Level-6 & 1.993~~\&~~0.558 &~~~~0.835,~~0.538~~\&~~0.458 &~0.433~~\&~~1.010\\
Level-7  & 2.036~~\&~~0.617 &~~~~0.940,~~0.603~~\&~~0.524 &~0.468~~\&~~1.050\\
\noalign{\smallskip}\hline
\end{tabular}
\end{table*}

It is also noticed that, although the unsigned  OMF in Case~1 is smaller than that in Case~2 at 1.01~$R_s$, they become nearly identical at 3~$R_s$ and 0.1~AU. Moreover, the ratio of open to total unsigned magnetic flux remains nearly the same for Cases~1 and~2 at both 1.01~$R_s$ and 3~$R_s$. This indicates that the higher-order spherical harmonic components retained in Case~2 increase both the open and closed magnetic flux in the low corona but do not affect the amount of OMF farther from the solar surface.
In addition, the OMF in Case~2 is consistently larger than that in Case~3 at 1.01~$R_s$, 3~$R_s$, and 0.1~AU, indicating that the evolution of the OMF is sensitive to heating source terms.

\begin{figure}[htpb]
\begin{center}
  \vspace*{0.01\textwidth}
    \includegraphics[width=0.9\linewidth,trim=1 1 1 1, clip]{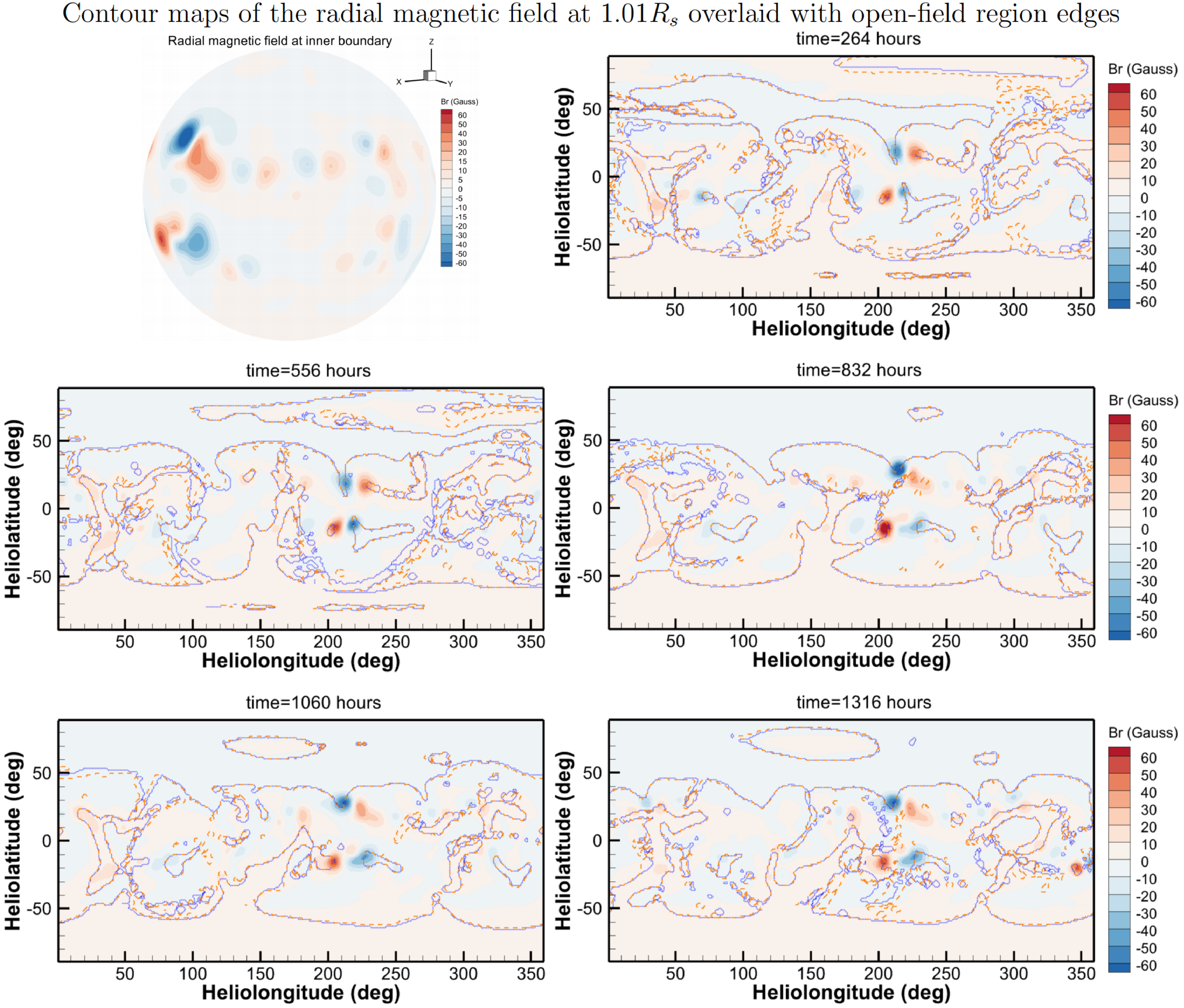}
\end{center}
\caption{Distributions of the radial magnetic field at 1.01~$R_s$ evaluated at the 264th, 556th, 832nd, 1060th, and 1316th hours for Case~3. The dashed orange, and dotted blue lines overlaid on the magnetic-field contours denote the boundaries of open-field regions derived from Cases~2, and 3, respectively. 
Evolution of the radial magnetic field at the inner boundary, as displayed in the top-left panel, during the simulated period is shown in online movie 1.
}\label{CHevolutionat1d01Rs}
\end{figure}
In Figure~\ref{CHevolutionat1d01Rs}, it is observed that a dipolar structure near $(\theta,\phi) = (20^{\circ}~\mathrm{N}, 220^{\circ})$ during the first 500~hours gradually shifts to $(\theta,\phi) = (20^{\circ}~\mathrm{N}, 220^{\circ})$ in the subsequent evolution. Meanwhile, a dipolar structure near $(\theta,\phi) = (10^{\circ}~\mathrm{S}, 210^{\circ})$ emerges at the 264th hour and persists throughout the subsequent evolution. At the 1316th hour, an additional dipolar structure appears near $(\theta,\phi) = (20^{\circ}~\mathrm{S}, 350^{\circ})$. 
Throughout the simulations, it is observed that in Cases~2 and 3, portions of both positive and negative poles of the dipolar structures are always overlain by open-field regions that pass through their centers. Moreover, the north and south polar regions remain dominated by open-field regions, with only minor differences between them in Cases~2 and~3.
The distributions of open-field regions in Cases~2 and~3 differ mainly in regions of weak magnetic field strength near the magnetic neutral lines (MNLs). 

\begin{figure}[htpb]
\begin{center}
  \vspace*{0.01\textwidth}
    \includegraphics[width=0.9\linewidth,trim=1 1 1 1, clip]{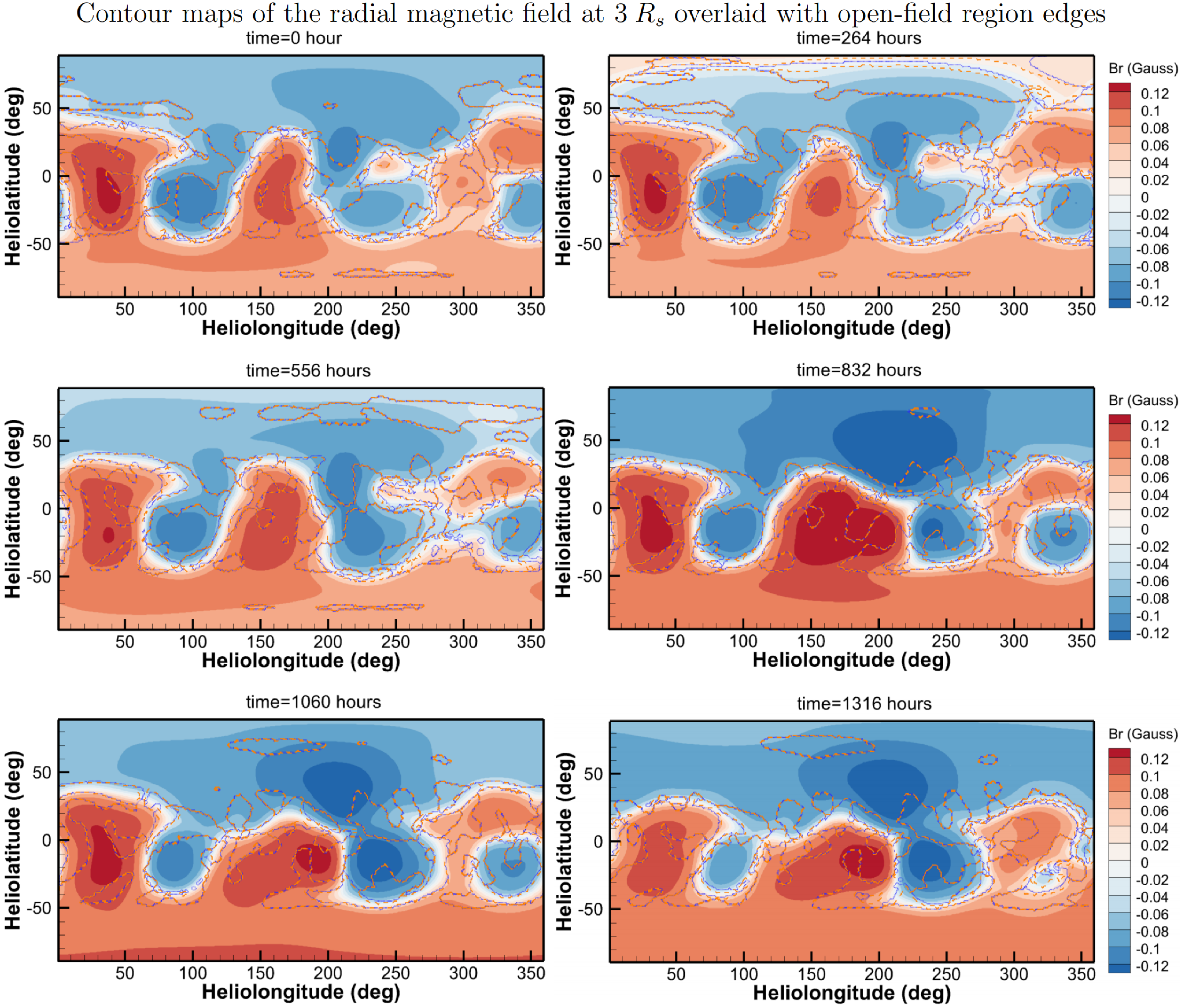}
\end{center}
\caption{Distributions of the radial magnetic field at 3~$R_s$ evaluated at the 0th, 264th, 556th, 832nd, 1060th, and 1316th hours for Case~3. The dashed orange, and dotted blue lines overlaid on the magnetic-field contours denote the boundaries of open-field regions derived from Cases~2, and 3, respectively.}\label{CHevolutionat3Rs}
\end{figure}
Figure~\ref{CHevolutionat3Rs} shows the evolution of the radial magnetic field strength at 3~$R_s$ from Case~3, together with the corresponding open-field region boundaries at 3~$R_s$ derived from Cases~2, and~3. It is observed that the maximum magnetic field strength decreases from approximately 70~G at 1.01~$R_s$ to around 0.13~G at 3~$R_s$. The variations in dipolar structures observed at 1.01~$R_s$ are also reflected at 3~$R_s$. For example, the enhancement of the magnetic field strength at the positive pole of the dipole near $(\theta,\phi) = (10^{\circ}~\mathrm{S}, 210^{\circ})$ at 1.01~$R_s$ around the 832nd hour corresponds to a patch of the obviously increased magnetic field centered around $(\theta,\phi) = (15^{\circ}~\mathrm{S}, 170^{\circ})$ at 3~$R_s$. 
Although the fraction of the spherical surface occupied by open-field regions increases significantly from 1.01~$R_s$ to 3~$R_s$, particularly for regions covering the northern and southern poles, a fraction of closed-field regions still persists at 3~$R_s$.
It is also noticed that the differences in the open-field region distributions between Cases~2 and~3 become less pronounced at 3~$R_s$ than at 1.01~$R_s$. 

For Case 1, the simulated open-field regions at 1.01 and 3 $R_s$ deviate more from those of Case 2 than do the open-field regions in Case 3. This demonstrates that although the small-scale magnetic structures in the low corona 
hardly affect the amount of simulated OMF far from the Sun, they significantly influence the distribution of open-field regions in the low corona.

\section{Concluding remarks}\label{sec:Conclusion}
Recently, a decomposed energy strategy \citep{Wang_COCONUT_DecE} that updates the internal and kinetic energy components separately at each time step, combined with an HLL Riemann solver that introduces an additional dissipation term in the energy equation, has been proposed and validated to significantly improve the numerical stability of the time-evolving coronal MHD model COCONUT \citep{wang2025COCONUTMayEvent,Wang2025_FirsttimeevolvingCOCONUT}. 
Combined with the high efficiency provided by implicit temporal integration, the time‑evolving coronal MHD model COCONUT is well suited for assessing uncertainties in solar‑maximum coronal simulations that involve low‑$\beta$ regions near active regions in the low corona. It enables exploration of the ``open-flux problem" during solar maximum using time-evolving MHD coronal simulations without discarding much of the small‑scale structure in the low corona, offering a more realistic representation than commonly used quasi‑steady approaches.

In this paper, we perform three time-evolving coronal simulations during CRs 2282 and 2283, driven by a sequence of hourly-updated photospheric magnetograms preprocessed with a 10th-order PF solver and a filtered 50th-order PF solver. The former removes many small-scale magnetic structures and reduces the magnetic field strength, thereby helping to avoid numerical stability issues at the cost of reduced accuracy. In contrast, the latter preserves much more small-scale magnetic structures and is accompanied by low-$\beta$ challenges. In addition, we adjust the empirically defined heating source terms to assess the impact of empirically defined heating uncertainties, one of the most challenging issues in coronal modeling, on the simulated evolution of magnetic flux at different heliocentric distances. 

These simulations reproduce reasonable high-density low-speed and low-density high-speed streamers. The temporal variation of the simulated low-latitude radial velocity is consistent with the temporal evolution of the corresponding central disk OMF. Additionally, the simulation results generally capture the observed white-light polarized brightness (pB) images and reflect the variations of solar wind structures inferred from interplanetary \textit{in situ} observations. To further improve the physical realism of the lower boundary, we implement non-uniform density boundary conditions based on parameterizations derived from local radiative MHD simulations performed with the \textit{Bifrost} code, thereby providing more physically informed conditions at the base of the corona, which is key for driving solar wind evolution. This parameterization should be regarded as a first proof of concept, and future work will focus on refining and extending it to further improve the realism of the lower boundary conditions. Further improvements, such as including physically based heating source terms like the wave turbulence-driven heating mechanism or more complex empirically defined volumetric heating terms that account for magnetic field topology, improving the spatial-temporal accuracy of magnetogram observations, especially in the polar regions, and incorporating transient events such as CMEs, may lead to better agreement between simulations and observations.

This paper reveals that the simulated unsigned OMF near the solar surface is comparable to the unsigned interplanetary OMF derived from WIND and PSP \textit{in situ} observations. However, it is noticed that the OMF can be reduced by up to $45\%$ from 1.01~$R_s$ to 0.1~AU in the MHD simulations. The decrease in the simulated unsigned OMF with increasing heliocentric distance occurs primarily in the low corona where numerous small-scale closed-field magnetic structures exist. Meanwhile, the closed magnetic field decreases more rapidly within 3~$R_s$, reducing from about $60\%$ of the total magnetic flux at 1.01~$R_s$ to less than $4\%$ at 3~$R_s$. 
In addition to interchange reconnection with closed magnetic loops \citep{Fisk2005,Iijima_2025}, some numerical issues can also account for the decrease in OMF with increasing heliocentric distance in the low corona. The low corona contains numerous small-scale closed magnetic structures embedded within the surrounding open field, with some of these closed magnetic structures exhibiting magnetic polarity opposite to that of the neighboring open field. This produces many polarity-inversion interfaces (PIIs) between open and closed magnetic fields. During the volume-integration procedure of the finite-volume method, magnetic fields of opposite polarity near the PII are partially canceled within a grid cell that passes through the PII, resulting in a reduction of the simulated OMF.
In contrast, the high corona is more potential and dominated by open fields, and only large structures, such as CMEs, which are very transient, streamers, pseudo-streamers, and cavities, can provide closed magnetic fields. 
Therefore, the decrease in unsigned OMF with increasing heliocentric distance is minor in the high corona.

The results also show that increasing the grid resolution by $2\times2$ in the tangential direction can increase the OMF by more than $10\%$ at different heliocentric distances, suggesting that coronal simulations incorporating adaptive mesh refinement \citep{Feng_2012,van_der_Holst_2014}, with finer grid resolution in regions in regions near PIIs, are likely to provide a promising approach to further reduce the magnetic-flux deficit while maintaining high computational efficiency.
This paper also shows that removing small-scale magnetic structures in the low corona, by preprocessing photospheric magnetograms with a PF solver that excludes high-order spherical harmonic components, reduces both the total and open unsigned magnetic fluxes in the low-coronal region, while having little impact on the simulated open unsigned flux beyond 3~$R_s$. 
Additionally, we find that preprocessing the photospheric magnetograms by the filtered PF solver \citep{MCCLARREN20105597, wang2025COCONUTMayEvent, Murteira2025} not only mitigates the ring artifacts in the magnetic field distribution caused by truncated spherical harmonic components \citep{Toth_2011}, but also reduces irregularly scattered closed-field fragments in the low corona, as shown in Appendix~\ref{Filterornot}.
Besides, the simulation results indicate that moderate adjustments to the empirically defined heating source term can effectively regulate the magnitude of the unsigned open magnetic flux, underscoring the need to adopt more realistic heating mechanisms in coronal MHD modeling.

It is noted that the simulated magnetic field strength at 0.1~AU, sampled by a virtual satellite placed at the same heliographic latitude as Earth, is approximately half of the hourly averaged WIND \textit{in situ} observations at 1~AU scaled by $\left( \frac{1~\mathrm{AU}}{0.1~\mathrm{AU}} \right)^2$. Consistently, the simulated unsigned magnetic flux at 0.1~AU is also about half of the unsigned magnetic flux inferred from the hourly averaged WIND interplanetary observations. This suggests that during solar maximum, although MNLs can undergo sharp deflections and even form isolated closed structures at 0.1~AU, even extending up to $50^{\circ}$ in latitude, the latitudinal variation of the magnetic field distribution can still be used to derive the total unsigned interplanetary magnetic flux. 
In future work, we will further investigate the ``open flux problem" by extending the quasi-realistic time-evolving MHD coronal simulations beyond 1~AU, enabling direct comparisons between simulation results and corresponding interplanetary \textit{in situ} observations during solar maximum.

Additionally, it shows that the unsigned OMF fluctuates significantly in solar-maximum CRs, with the maximum unsigned open flux being more than 1.4 times the minimum during CRs 2282 and 2283. This is due to the relatively rapid evolution of magnetic flux emergence and cancelation during solar-maximum CR periods, as indicated in online movie~1. It produces significant variations in the unsigned magnetic flux observed in magnetograms, leading to the fluctuations observed in the open flux. The pronounced variations in the shape and amplitude of the periodically recurring peaks and troughs in the temporal evolution of the central-disk OMF, as viewed from Earth, also indicate substantial variability of the OMF within a single solar-maximum CR. This underscores the need to evaluate magnetic flux evolution during solar maximum using time-evolving coronal simulations driven by high-cadence, synchronized magnetograms rather than the commonly employed quasi-steady-state simulations constrained by static synoptic magnetograms.
Recently, \cite{Downs2025PSI} has also performed time-evolving coronal simulations around CR~2282 using another state-of-the-art MHD coronal model, the Magnetohydrodynamic Algorithm outside a Sphere \citep[MAS;][]{Lionello_2023,Mason_2023}. 
As discussed in \cite{Baratashvili2025}, MAS employs more realistic synchronized magnetograms, in which the magnetic fields at different longitudes correspond to the same time, to drive coronal evolution. MAS also includes the transition region and adopts a more physical consistent wave–turbulence–driven heating mechanism. We will incorporate these more realistic settings into the time-evolving coronal model COCONUT, which employs the more efficient implicit temporal integration method and the decomposed energy strategy to address low-$\beta$ issues, in future work.

\textbf{Acknowledgments}:
This project has received funding from the European Research Council Executive Agency (ERCEA) under the ERC-AdG agreement No.\ 101141362 (Open SESAME). 
These results were also obtained in the framework of the projects FA9550-18-1-0093 (AFOSR), C16/24/010  (C1 project Internal Funds KU Leuven), G0B5823N and G002523N (WEAVE) (FWO-Vlaanderen), and 4000145223 (SIDC Data Exploitation (SIDEX), ESA Prodex).
This work is also supported by the BK21 FOUR program of the Graduate School, Kyung Hee University (GSX-20242364 and GSX-20253142).
The resources and services used in this work were provided by the VSC (Flemish Supercomputer Centre), funded by the Research Foundation – Flanders (FWO) and the Flemish Government. This work utilises data obtained by the Global Oscillation Network Group (GONG) program, managed by the National Solar Observatory and operated by AURA, Inc., under a cooperative agreement with the National Science Foundation. The data were acquired by instruments operated by the Big Bear Solar Observatory, High Altitude Observatory, Learmonth Solar Observatory, Udaipur Solar Observatory, Instituto de Astrof{\'i}sica de Canarias, and Cerro Tololo Inter-American Observatory. The authors also acknowledge the use of the STEREO/SECCHI data produced by a consortium of the NRL (US), LMSAL (US), NASA/GSFC (US), RAL (UK), UBHAM (UK), MPS (Germany), CSL (Belgium), IOTA (France), and IAS (France), and use of NASA/GSFC's Space
Physics Data Facility's OMNIWeb service. 

\begin{appendix}
\section{Coronal evolutions under different numerical setups}\label{sec:EvolutionofCoronalStructure}

\begin{figure}[htpb]
\begin{center}
  \vspace*{0.01\textwidth}
    \includegraphics[width=0.8\linewidth,trim=1 1 1 1, clip]{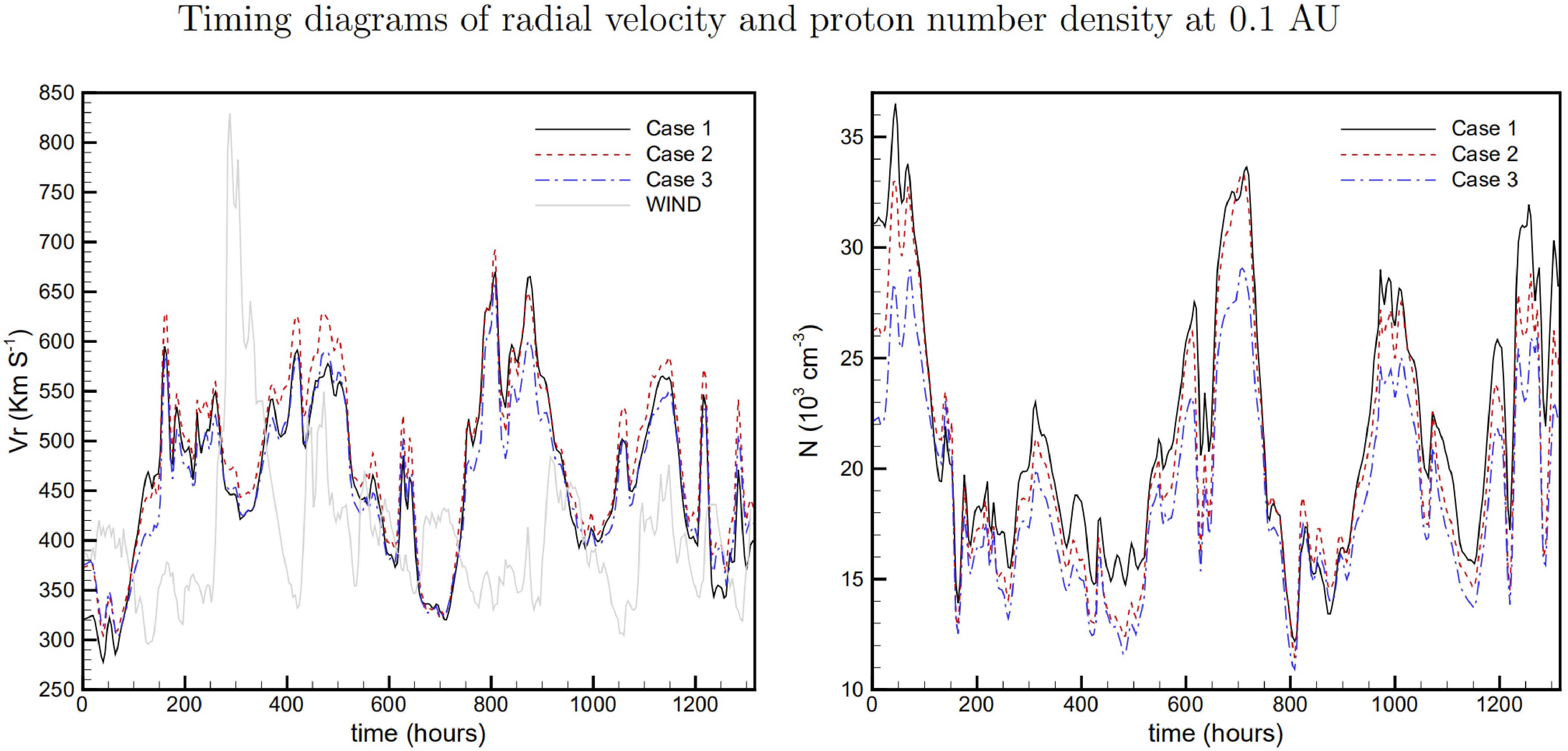}
\end{center}
\renewcommand{\thefigure}{A.~1}
\caption{
Timing diagrams of the radial velocity $V_r$ ($\rm km~s^{-1}$; left) and proton number density ($\rm 10^3 ~ cm^{-3}$; right) measured by a virtual satellite located at 21.5~$R_s$. The virtual satellite is positioned at the same latitude as Earth but lags by $60^\circ$ in longitude. The solid black, dashed red, and dashed-dot blue lines represent the time-evolving simulation results from Cases~1,~2, and~3, respectively, while the solid gray line shows the radial velocity (left) observed by the WIND spacecraft.}\label{1DTimingDiagram}
\end{figure}
During these simulations, a virtual satellite is placed at 21.5 $R_s$. It lags the Sun–Earth line by $60^{\circ}$, assuming that the solar wind takes approximately 100 hours to travel from 0.1 to 1 AU \citep{wang2025COCONUTMayEvent,wang2025sipifvmtimeevolvingcoronalmodel}. Figure~\ref{1DTimingDiagram} shows that the simulated radial velocity in Case~2 is generally higher than in Cases~1 and~3, while the density in Case~1 is typically higher than in Cases~2 and~3, with the most pronounced differences occurring near the peaks. 
The profiles of these variables are largely consistent across the three cases, the comparison between Cases~1 and~2 shows that the magnetogram over-filtered by the 10th-order PF solver leads to relatively higher plasma density and lower radial velocity, whereas the comparison between Cases~2 and~3 indicates that a minor modification in the radial distribution of the heating source term can reduce both the radial velocity and plasma density, with an effect that is more pronounced than that resulting from variations in magnetogram filtering. This underscores the need to explore more realistic heating mechanisms for coronal MHD simulations.

It is also noted that the simulated radial-velocity trough, centered around the 50th hour, occurs approximately 70 hours later than the observed one. The velocity peak and trough, centered around the 300th and 800th hours, are mismatched in the simulations, with a trough and a peak. Meanwhile, the simulated radial velocity between the 450th and 650th hours is largely consistent with the observations, and the simulated troughs around the 1000th, 1200th, and 1250th hours closely match the observed ones. 
The oversimplified empirically defined heating source terms, the lack of accurate magnetic field observations in the polar regions, the exclusion of transient events such as CMEs from the simulations, and the neglect of the complex evolution of solar-wind structures in interplanetary space can all contribute to the discrepancies between simulations and \textit{in situ} observations. 

\begin{figure}[htpb]
\begin{center}
  \vspace*{0.01\textwidth}
    \includegraphics[width=0.8\linewidth,trim=1 1 1 1, clip]{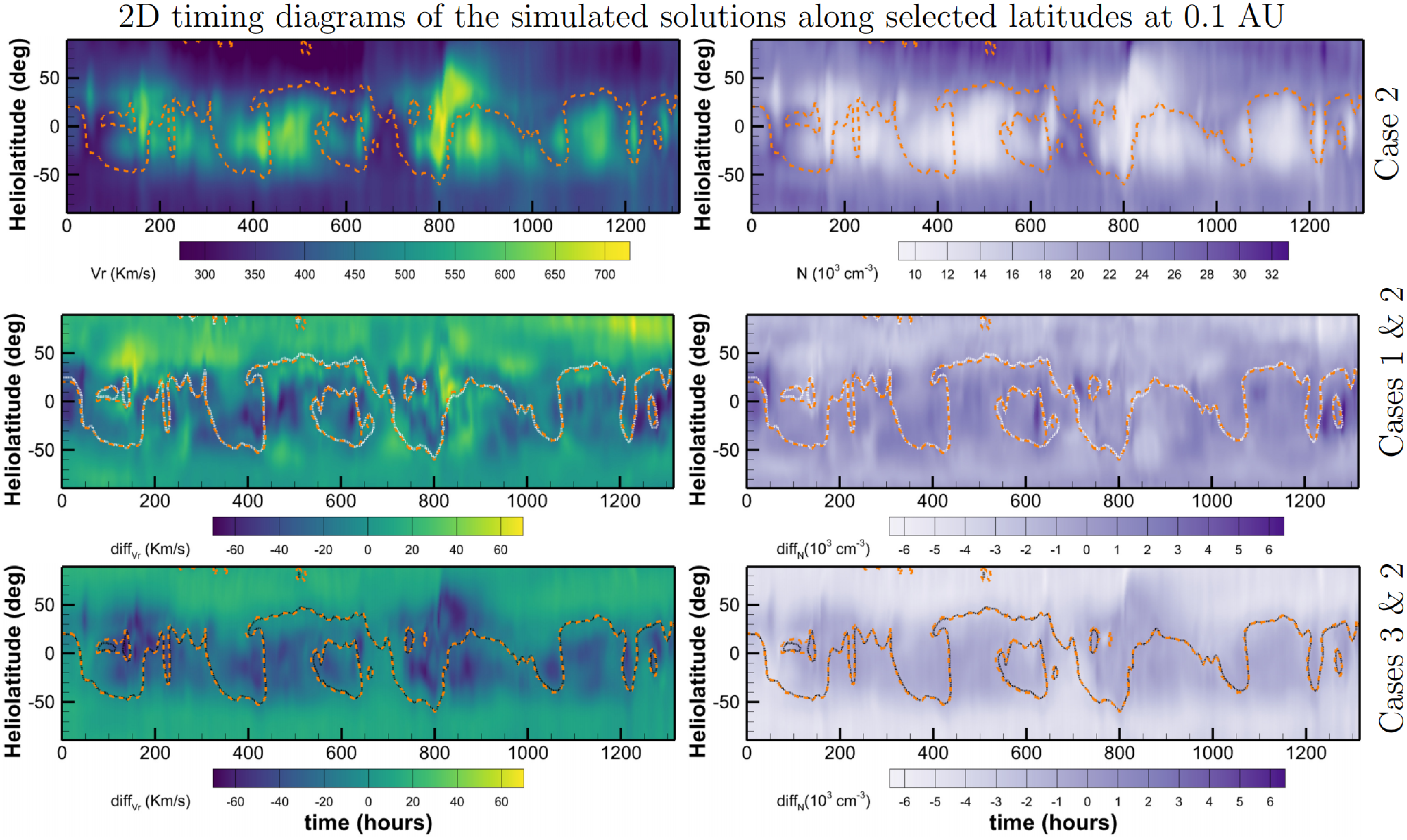}
\end{center}
\renewcommand{\thefigure}{A.~2}
\caption{
Timing diagrams of the simulated radial velocity $V_r$ ($\rm km\,s^{-1}$; left) and plasma number density ($10^3\,\rm cm^{-3}$; right) from Case~2 (top) at 0.1~AU, evaluated along the latitudes intersected by longitude $0^\circ$. The middle and bottom panels show the differences in radial velocity and density between Cases 1 and 2, and between Cases 3 and 2, respectively, at the same locations as the top panels.
The dashed white, dotted orange, and dotted black lines denote the MNLs derived from Cases~1, 2, and 3, respectively.}\label{2DTimingDiagram}
\end{figure}
Figure~\ref{2DTimingDiagram} shows that between 300 and 600 hours the northern polar region is dominated by low-velocity, high-density flows. This is associated with the appearance of closed magnetic field structures in the northern polar region, as illustrated in Figure~\ref{CHevolutionat1d01Rs}. Additionally, the MNLs in the three cases are consistent with each other. In Case~1, the removed small-scale magnetic structures in the low corona lead to lower velocities and higher densities near the MNLs, but to higher velocities and lower densities in regions farther from the MNLs. This indicates that although small-scale magnetic structures in the low corona have a minor effect on the total amount of open flux far from the Sun, they can significantly affect the latitudinal distribution of open-field regions and consequently influence the distribution of solar wind streamers. Furthermore, the area-averaged radial velocity $V_{r,\rm ave}$ and proton number density $\rm N_{\rm ave}$ evaluated over the panels in Figure~\ref{2DTimingDiagram} are listed below.
 \begin{enumerate}
     \item Case~1: $V_{r,\rm ave}=433.21~\rm km\,s^{-1}$ and $\rm N_{\rm ave}=20.11\times10^3\,\rm cm^{-3}$.
     \item Case~2: $V_{r,\rm ave}=436.82~\rm km\,s^{-1}$ and $\rm N_{\rm ave}=19.92\times10^3\,\rm cm^{-3}$.
     \item Case~3: $V_{r,\rm ave}=426.73~\rm km\,s^{-1}$ and $\rm N_{\rm ave}=17.53\times10^3\,\rm cm^{-3}$.
\end{enumerate} 
These results demonstrate that the minor adjustments to the heating distribution introduced in this paper regulate the magnitudes of the simulated velocity and density at 0.1~AU more effectively than the truncation of high-order spherical harmonic components during magnetogram preprocessing, while all three cases preserve the overall tangential coronal structures.

\begin{figure*}[htpb]
\begin{center}
  \vspace*{0.01\textwidth}
    \includegraphics[width=0.8\linewidth,trim=1 1 1 1, clip]{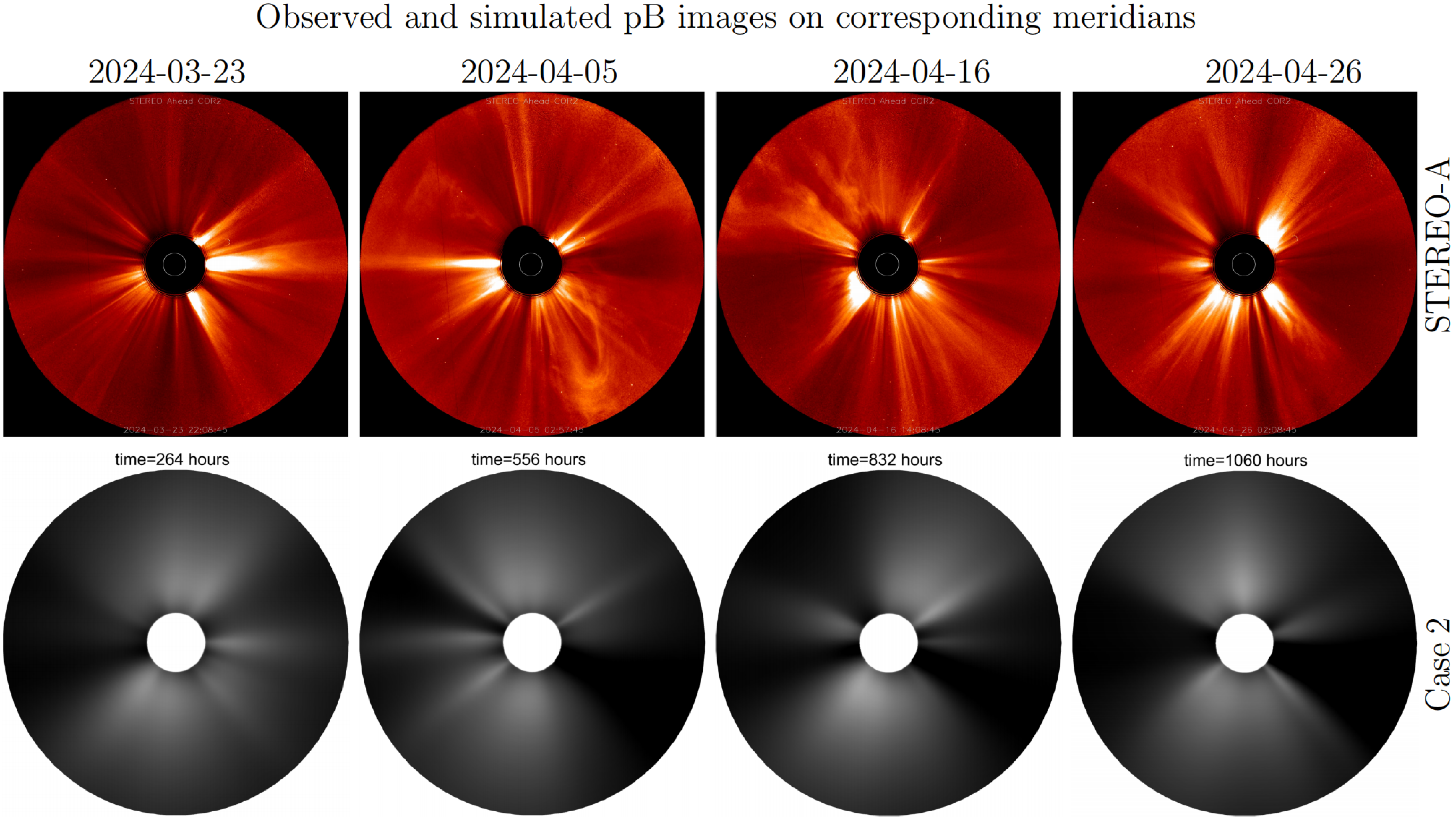}
\end{center}
\renewcommand{\thefigure}{A.~3}
\caption{
White-light pB images observed by COR2/STEREO-A (first row) and synthesised from the coronal simulation results of Case~2 (second row). The evolution
of these simulated images during the simulated period is shown in online movie 2.}\label{pBMeridian}
\end{figure*} 
The simulated pB images \citep{Wang_COCONUT_DecE} in Cases~1,~2, and~3 agree well with each other. Figure~\ref{pBMeridian} shows that the simulation at the 264th hour captures four bright structures, and the observed structure centred near $30^{\circ}$~N on the east limb is shifted northward by approximately $20^{\circ}$ in the simulation.
In addition, the simulated bright structures near both polar regions are absent from the observations. At the 556th hour, the simulations reproduce the observed features at low and middle latitudes. Meanwhile, some bright structures still appear near the polar regions, whereas they are absent in the observations.  
At the 832nd hour, the observed bright structures around $50^{\circ}$~S on both the west and east limbs merge into a wide bright feature covering the southern polar region.
Meanwhile, the observed bright structure near the northern pole is rotated clockwise by approximately $20^{\circ}$, and the two simulated low-latitude bright structures agree well with the observations. At the 1060th hour, the high-latitude bright structure covering the southern pole is successfully reproduced. However, the bright structures along the equator on the east limb and near $60^{\circ}$ on the west limb are absent in the simulations, whereas the observed bright structures near the northern pole and around $50^{\circ}$ N in the east limb are basically reproduced in simulations. From these comparisons, it is evident that the pB images in the polar regions show the largest discrepancies between the simulations and observations, indicating an urgent need to drive time-evolving coronal evolution during solar maximum with more realistic magnetograms and improved observations of the polar regions. The comparison with pB observation is intended only to demonstrate that our model can qualitatively reproduce the observations. In future work, a more quantitative validation using a pB comparison will be conducted, where COCONUT will incorporate more realistic synchronised magnetograms and a more self-consistent wave–turbulence–driven heating mechanism.

\section{ Impact of filtered and unfiltered PF solver preprocessing of initial magnetograms}\label{Filterornot}
\begin{figure}[htpb]
\begin{center}
  \vspace*{0.01\textwidth}
    \includegraphics[width=0.8\linewidth,trim=1 1 1 1, clip]{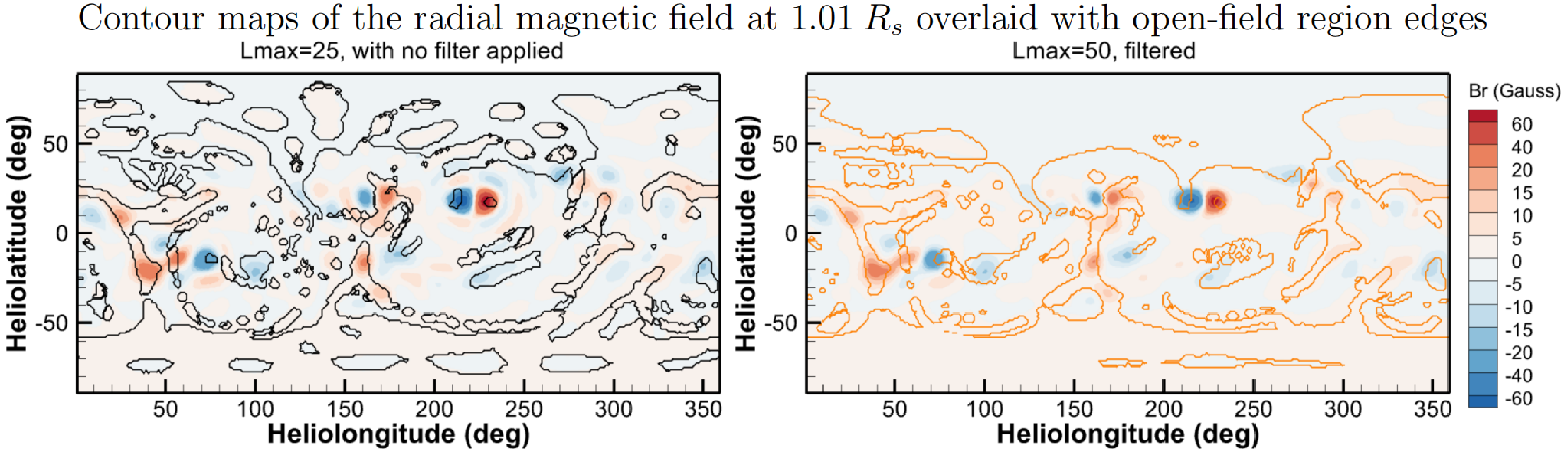}
\end{center}
\renewcommand{\thefigure}{B.~1}
\caption{Distributions of the radial magnetic field at 1.01~$R_s$ preprocessed using the 25th-order PF solver (left) and the 50th-order filtered PF solver as described in Eq.~\ref{filteredspherical} (right). These results correspond to the 0th hour of the simulated period. The solid lines overlaid on the magnetic-field contours indicate the boundaries of open-field regions derived from the corresponding coronal simulation results.
}\label{CH25VSFiltered50}
\end{figure}
Figure~\ref{CH25VSFiltered50} shows the distributions of the radial magnetic field at 1.01~$R_s$, preprocessed using a 25th-order PF solver without filtering the spherical harmonic functions (left, Case~4) and a 50th-order PF solver with filtered spherical harmonic functions (right, Case~3), together with the corresponding open-field regions. These images are derived from the simulated results at the 0th hour. The ring-like magnetic field structures evident in the left panel disappear in the right panel. The closed-field patches scattered within the polar regions are significantly reduced in the right panel. The open magnetic flux, normalized in the same way as in Figure~\ref{Fluxevolution}, is 0.939 and 0.835 at 1.01~$R_s$, 0.531 and 0.538 at 3~$R_s$, and 0.453 and 0.458 at 0.1~AU, for Cases~4 and 3, respectively. Correspondingly, the ratios of open to total unsigned flux are 0.379 and 0.419 at 1.01~$R_s$, and 0.962 and 0.964 at 3~$R_s$, for Cases~4 and 3, respectively. These results further demonstrate that the spherical harmonic preprocessing of the magnetograms, which removes several small-scale magnetic structures, has little impact on the amount of simulated open unsigned flux at larger heliocentric distances, although it does influence the distribution of open-field regions and the amount of open unsigned flux in the low corona.

\end{appendix}
\pagebreak

\bibliographystyle{aasjournal}
\bibliography{SIPandCOCONUT}

\end{document}